\documentclass[12pt,preprint]{aastex}

\newcommand{\eg}{e.g.,\ }

\newcommand{\Msun}{M_{\odot}}

\newcommand{\kms}{km~s$^{-1}$}

\newcommand{\OI}{O~{\sc i}}

\newcommand{\MgI}{Mg~{\sc i}}
\newcommand{\SiI}{Si~{\sc i}}
\newcommand{\SiII}{Si~{\sc ii}}

\newcommand{\SI}{S~{\sc i}}
\newcommand{\CaII}{Ca~{\sc ii}}

\newcommand{\FeII}{Fe~{\sc ii}}
\newcommand{\FeIII}{Fe~{\sc iii}}

\newcommand{\Fefs}{$^{56}$Fe}
\newcommand{\Cofs}{$^{56}$Co}
\newcommand{\Nifs}{$^{56}$Ni}
\newcommand{\Mej}{$M_{\rm ej}$}
\newcommand{\KE}{$E_{\rm kin}$}

\begin{document}

\title{The Aspherical Properties of the Energetic Type Ic SN\,2002ap 
	as Inferred from its Nebular Spectra$^1$}

\author{P. A. Mazzali\altaffilmark{2,3,4,5},
K.~S.~Kawabata\altaffilmark{6}, 
K.~Maeda\altaffilmark{2,7}, 
R.~J.~Foley\altaffilmark{8}, 
K.~Nomoto\altaffilmark{4,5},
J.~Deng\altaffilmark{9},
T.~Suzuki\altaffilmark{4}, 
M.~Iye\altaffilmark{10,11},
N.~Kashikawa\altaffilmark{10},
Y.~Ohyama\altaffilmark{12},
A.~V. Filippenko\altaffilmark{8},
Y.~Qiu\altaffilmark{9}, and
J.~Wei\altaffilmark{9}}

\altaffiltext{1}{Based in part on data obtained at the Subaru Telescope,
 which is operated by the National Astronomical Observatory of
 Japan (NAOJ). Also based in part on data obtained at the University
of California's Lick Observatory.}
\altaffiltext{2}{Max-Planck Institut f\"ur Astrophysik, 
  	Karl-Schwarzschild-Str.\ 1, 85748 Garching, Germany.} 
\altaffiltext{3}{National Institute for Astrophysics--OATs, Via G.B. Tiepolo, 11,
	34143 Trieste, Italy.}
\altaffiltext{4}{Department of Astronomy, School of Science, 
	University of Tokyo, Bunkyo-ku, Tokyo 113-0033, Japan.}
\altaffiltext{5}{Research Center for the Early Universe, School of Science, 
	University of Tokyo, Bunkyo-ku, Tokyo 113-0033, Japan.}
\altaffiltext{6}{Hiroshima Astrophysical Science Center,
        Hiroshima University, Hiroshima 739-8526, Japan.}
\altaffiltext{7}{Department of Earth Science and Astronomy,
         College of Arts and Sciences, University of Tokyo, Meguro-ku,
         Tokyo 153-8902, Japan.}
\altaffiltext{8}{Department of Astronomy, University of California, Berkeley, 
	CA 94720-3411.}
\altaffiltext{9}{National Astronomical Observatories, CAS, 20A Datun Road,
	Chaoyang District, Beijing 100012, China.}
\altaffiltext{10}{National Astronomical Observatory of Japan, Mitaka, Tokyo
	181-8588, Japan.}
\altaffiltext{11}{Department of Astronomical Science, Graduate University for
	Advanced Studies, Mitaka, Tokyo 181-8588, Japan.}
\altaffiltext{12}{Subaru Telescope, National Astronomical Observatory of Japan, 
	650 North A'ohoku Place, Hilo, HW 96720.}
	
\begin{abstract}

The nebular spectra of the broad-lined, SN\,1998bw-like Type Ic SN\,2002ap are
studied by means of synthetic spectra. Two different modelling techniques are
employed. In one technique, the SN ejecta are treated as a single zone, while in
the other a density and abundance distribution in velocity is used from an
explosion model. In both cases, heating caused by $\gamma$-ray and positron
deposition is computed (in the latter case using a Monte Carlo technique to
describe the propagation of $\gamma$-rays and positrons), as is cooling via
forbidden-line emission. The results are compared, and although general agreement
is found, the stratified models are shown to reproduce the observed line profiles
much more accurately than the single-zone model.  The explosion produced $\sim
0.1\, \Msun$ of \Nifs.  The distribution in velocity of the various elements is in
agreement with that obtained from the early-time models, which indicated an
ejected mass of $\sim 2.5\, \Msun$ with a kinetic energy of $4 \times
10^{51}$\,erg. Nebular spectroscopy confirms that most of the ejected mass ($\sim
1.2\, \Msun$) was oxygen. The presence of an oxygen-rich inner core, combined with
that of \Nifs\ at high velocities as deduced from early-time models, suggests that
the explosion was asymmetric, especially in the inner part.

\end{abstract}


\section{Introduction}

One of the most exciting recent discoveries in astrophysics is that the
nearest  observed long-duration gamma-ray bursts (GRBs) are associated with
supernovae  (SNe) (\eg GRB980425/SN\,1998bw, \citealt{gal98};
GRB030329/SN\,2003dh, \citealt{sta03}; GRB031203/SN\,2003lw, \citealt{mal04}).
Adding to the excitement, it turned out that these SNe are not at all ordinary.
Spectroscopically, they are of Type Ic (no H, no He, weak Si lines; see
\citealt{fil97} for a review of SN classification), but their distinguishing 
feature is that unlike normal SNe~Ic, they show extremely broad absorption
lines dominated by Fe, Ca, and O. Very high expansion velocities, reaching $0.1
c$, are easily observed at early times. 

Broad lines suggest a very energetic explosion, or at least a large amount of
energy per unit mass. Estimates of the explosion kinetic energy assuming
spherical symmetry for SNe 1998bw, 2003dh, and 2003lw are \KE\,$\approx (3 - 6)
\times 10^{52}$\,erg (\eg \citealt{iwa98,nak01,maz03, den04,maz06a}). This is
more than one order of magnitude larger than in typical core-collapse SNe,
which have \KE\,$\approx 10^{51}$\,erg. Accordingly, these exceptionally
powerful SNe have sometimes been called ``hypernovae'' (\eg \citealt{nom05}).
In the presence of significant deviations from spherical symmetry, these
estimates may be reduced, but even then the energies remain large (\eg
\citealt{mae06}). The mechanism behind these events is thought to be the
collapse of the stripped core of a massive star ($\sim 25 - 60\, \Msun$) to a
black hole \citep{mac99}.

Additionally, evidence has been obtained that X-ray flashes (XRFs), the soft
analogues of GRBs \citep{heise01}, are also linked to SNe
\citep[e.g.,][]{fyn03,mod06,pian06}. The SNe associated with XRFs also seem to
be overenergetic, but not as much so as GRB-SNe \citep{tom03,maz06b}. They have
been suggested to be the result of the collapse of the stripped carbon-oxygen
core of stars of originally $\sim 20\, \Msun$ to a neutron star
\citep{maz06b,mae07}. If the neutron star is born highly magnetized and rapidly
spinning --- a ``magnetar'' \citep{dun92,tho04} --- it may cause the explosion
to be overenergetic and produce the XRF \citep{maz06b,mae07}. 

One difference between SNe in GRBs and XRFs may be the degree of asphericity.
GRB-SNe are thought to be highly aspherical, as GRBs are generally believed to
be highly asymmetric phenomena (see \citealt{woo06} for a recent review of the
SN-GRB connection). For XRF-SNe, the degree of asphericity may be smaller
\citep{maz07}.

Evidence for asphericity is not easy to glean from early-time data.   The
significant mixing outward of \Nifs\ required to reproduce the early rise of the
light curve is a general feature of hypernovae \citep{mae03} and suggests an
aspherical explosion, as does the connection of these SNe with GRBs. The best time
to look for signatures of asphericity, however, is starting a few months after the
explosion, when the SN becomes nebular, exposing the deepest parts of the ejecta.
In that phase, both the spectrum and the light curve have a characteristic
behavior. The light curve of SN\,1998bw showed a phase of strictly exponential
decline, but at a rate steeper than that of \Cofs\ \citep{pat01}. In this phase
the luminosity is larger than what is predicted by spherically symmetric models
that fit the peak of the light curve \citep{nak01}. \citet{mae03} showed that this
behavior, which is actually  not unusual in SNe~Ic, can be explained using a
modified spherically symmetric model where a relatively massive but slowly
expanding inner core is placed in the center of the expanding ejecta. Although
this is technically still a spherically symmetric model, one-dimensional explosion
simulations do not predict such a density distribution, which suggests that we are
observing an aspherical explosion, similar perhaps to the collapsar model of
\citet{mac99}.

The nebular spectra provide more direct evidence. In SN\,1998bw  the [\FeII] lines
are broader than the [\OI] $\lambda\lambda$6300, 6363 doublet \citep{maz01}.  This
is also something that cannot be explained in the context of a spherically
symmetric model of the collapse and explosion of a CO core, as in such a model
\Nifs\ (which decays into \Cofs\, and then into \Fefs) is synthesized near the
compact remnant and therefore is always located at smaller velocities than O,
which is left from the progenitor in the unprocessed outer layers. \citet{mae02}
showed that this configuration can be obtained in an axisymmetric explosion. In
such an explosion, most of the \KE\ is released along  the ``jet'' axis, which is
probably linked to the launching of the GRB, and \Nifs\ is synthesized along that
direction and ejected at a high velocity. Away from the jet axis, however, less
kinetic energy is deposited and burning proceeds much less efficiently. Therefore,
in these directions large amounts of unburned O are ejected at low velocity. For a
near-polar viewing angle, this scenario naturally leads to broad [\FeII] lines and
a narrow [\OI] line at late times. For an equatorial view, on the other hand, the
[\OI] line should have a characteristic double-peaked profile \citep{mae02}, as
was indeed observed in the SN~Ic 2003jd \citep{maz05}. Further evidence that
SNe~Ic are aspherical also comes from their high degree of polarization
\citep{wang01,leo02,fil04,leo05}.

For XRF/SNe, the case for asphericity is much weaker. The early-time spectra of
SN\,2006aj did not show very broad features \citep{pian06}. Radio observations
\citep{sod06} suggest a very broad opening angle ($> 60^{\circ}$). The broad
nebular lines are not inconsistent with spherically symmetric ejecta
\citep{maz07}. Nevertheless, the presence of emission at velocities below
2000\,\kms\ is at odds with the prediction of one-dimensional explosion models,
that  a density hole is present at the lowest velocities \citep{maz07,mae07}.

Besides those clearly associated with GRBs or XRFs, other energetic SNe Ic have
been observed. Energetic SNe~Ic, with or without a GRB, are recognized from the
extreme width (up to $\sim 0.1 c$) of their spectral lines at early times, and 
have sometimes also been called hypernovae (although their total energy is not
always much higher than that of normal SNe~Ic). The nearest such object ever
observed was SN\,2002ap in M74. SN\,2002ap was immediately recognized as a
broad-lined event, similar spectroscopically to SN\,1998bw \citep{kinu02}, and
was therefore intensively observed. SN\,2002ap remained much less luminous
than SN\,1998bw \citep{maz02,galyam02}, and it was not significantly more
luminous than the average SN~Ic. Also, SN\,2002ap was not seen in association
with a GRB \citep{hurley02,galyam02}. 

\citet{maz02} modelled the early-time light curve and spectral evolution of
SN\,2002ap and derived values of the ejected mass (\Mej\,$= 2.5-5\, \Msun$), the
kinetic energy of the explosion (\KE\,$= 4-10 \times 10^{51}$\,erg), and the
mass of \Nifs\ synthesised in the explosion ($M$(\Nifs)\,$=0.07\Msun$). While
the mass of \Nifs\ is similar to that of ``normal'' SNe~Ic \citep{sauer06}, and
much smaller than that of hypernovae linked to GRBs, both \Mej\ and \KE\ are
intermediate between those of normal SNe~Ib/c and GRB-SNe. The uncertainty in
the values of \Mej\ and \KE\ follows from assuming either the absence of any He
envelope (lower bound) or the presence of a maximal, $2.5 \Msun$ helium envelope,
the presence of which would not affect the light curve significantly but is at
the same time not supported by the spectral appearance of SN\,2002ap. 
Interestingly, the light-curve behavior at advanced phases was very similar to
that of SN\,1998bw \citep{tom05}, suggesting a common nature for the two
events. 

\citet{maz02} argued that SN\,2002ap was the $\sim 5~\Msun$ carbon-oxygen core
of a star of initially $\sim 25~\Msun$ that collapsed to a black hole. On the
other hand, no GRB was detected, only weak radio emission was detected from the
SN \citep{berger02}, and the X-ray signal was very weak \citep{soria03},
suggesting that little or no relativistic ejecta were produced. This may be
related to the relatively small mass of the collapsing star, which may have
been close to the minimum required to form a black hole.  It is therefore
interesting to explore further the nature of this object. As in the case of
SNe\,1998bw and SN\,2006aj, this can be done by modelling the nebular spectra.

\citet{fol03} published a series of late-time spectra of SN\,2002ap, while
\citet{kaw02} published a single spectrum.  The nebular phase in SN\,2002ap
started to develop rather early, as in SN\,1998bw but unlike other hypernovae
such as SN\,1997ef \citep{maz04}. At an age of $\sim 4$ months, the SN spectra
were already fully nebular \citep{fol03}. The optical spectra were dominated by
very strong [\OI] $\lambda\lambda$6300, 6363 emission, and showed strong \CaII\
emission lines, similar to SN\,1998bw. Unlike the case of SN\,1998bw, however, the
[\FeII]-dominated blend near 5200~\AA\ is rather weak. This is not surprising,
since SN\,1998bw is thought to have produced $\sim 4$ times as much \Nifs\ as
SN\,2002ap \citep{maz06b}. Still, the [\FeII] lines in SN\,2002ap are
sufficiently pronounced that modelling can be meaningfully attempted. One
peculiarity of SN\,2002ap is the great strength of the \MgI] $\lambda\lambda$4571
line, which may suggest that the envelope of SN\,2002ap was more thoroughly
stripped than that of SN\,1998bw. This would favor the lower bound of the mass
and energy estimates of \citet{maz02}, which were obtained assuming the absence
of any significant He layer.

\section{The Spectral Dataset}

In this paper, we model all the spectra published by \citet{fol03}.
Additionally, we present and model three additional late-time spectra.
Two of these spectra were obtained with the Subaru telescope.
The other spectrum was obtained at the National Astronomical Observatory of China
(NAOC, formerly Beijing Astronomical Observatory, BAO).

The Subaru spectra were obtained on 2002 Jun 7.6 (UT dates are used 
throughout this paper) and Sep 15.6 with FOCAS.
For both spectra, the exposure time was 240 s, and grism B300 and Y47 order-cut
filters were used. A $0\farcs 8$ wide slit was used for the June spectrum and
the resulting spectral resolution ($\lambda/\Delta\lambda$) is $\sim 650$. For
the September spectrum, a $2\farcs 0$ slit was used, but the spectral resolution 
is still similar to that of the June spectrum because the seeing size was
$0\farcs 7$--$0\farcs 8$. The flux was calibrated using observations of
either Feige 110 or G191B2B obtained on the same night as the target spectra.

The BAO spectrum was obtained on 2002 July 11.8 with the NAOC
2.16~m telescope at Xinlong Observatory (then BAO).
The observations were carried out with an OMR (Optomechanics Research, Inc.)
spectrograph, using a Tektronix $1024\times 1024$ pixel CCD as the detector.
A grating of 300 g mm$^{-1}$ was used, which provided a spectral resolution
of $\sim 10-11$~\AA. The exposure time was $\sim 40$ min.

The newly published spectra are shown in Figure 1. 
The NAOC spectrum is reasonably consistent with the almost contemporaneous Lick
spectrum published by \citet{fol03}, and therefore it was modelled but is not 
explicitly shown in the Figures in the following sections. 

The 11 spectra we used for modelling cover a time span of almost 9 months, from
June 2002 through February 2003, corresponding to SN ages of 4--13 months given
that the SN exploded on 29 January 2002 \citep{maz02}.  Unfortunately, the
wavelength coverage is not uniform. In particular, the Subaru spectra extend over
a shorter range (4700--9000\,\AA) than the other spectra, missing the important
\MgI] line. Models based on the Subaru spectra are therefore somewhat less
reliable than those based on the Lick spectra. All spectra were 
calibrated by using the available photometry
\citep{fol03} with the exception of the last three, for which an extrapolation of
the light curve was used.

\section{Modelling Technique}

In order to model the nebular spectra of SN\,2002ap, we used our non-local
thermodynamic equilibrium (non-LTE) code \citep{maz01}.  The code computes the
heating of the gas following the deposition of $\gamma$-rays and positrons emitted
by the decay of \Cofs\ into \Fefs.  Heating is balanced by cooling via nebular
line emission. The emission rate in each line is computed solving a non-LTE matrix
of level populations \citep{axe80}.  In the original version \citep{RLL92}, a
homologously expanding nebula of finite extent, uniform density, and uniform
composition is assumed, and the emission spectrum is obtained assigning to all
lines a parabolic profile, bounded by the velocity of the outer edge of the
nebula.

Together with this classical version, a more advanced --- although more
model-dependent --- version has been developed and is used in this work. In
this new version, stratification in density and abundance is adopted. The
density profile is taken from explosion models, and $\gamma$-rays and positrons
are emitted at various depths according to the distribution of \Nifs. Their
propagation and deposition is followed using a Monte Carlo scheme similar to
that discussed by \citet{cap97} for their light-curve models. A constant
$\gamma$-ray opacity ($\kappa_{\gamma} = 0.027$\,cm$^2$\,g$^{-1}$) and a
constant positron opacity ($\kappa_{e^+} = 7$\,cm$^2$\,g$^{-1}$) are assumed.
The heating and cooling of the gas are then computed in non-LTE in each radial
shell, and so is the line emissivity. The line profiles from each shell are
assumed to be truncated parabolas, the inner truncation point corresponding to
the inner boundary of the shell considered. These truncated parabolas are then
summed to produce the emerging spectrum. Line profiles thus depend on the
density and abundance distributions. Therefore, this approach constitutes a
test of the explosion models which are used to simulate the light curves and
the early-time spectra, especially for what concerns the innermost part, which
is only visible directly in the nebular phase.

Because the stratified code depends more directly on the adopted explosion
model  than the one-zone version, it is useful to compare the results of the
two approaches. The next two sections deal with both cases in turn.

One important ingredient for the modelling is the assumed distance and reddening
to the SN. In the case of SN\,2002ap, the distance is highly uncertain, as
discussed by \citet{vinko04}. \citet{sha96}, using the brightest blue
supergiant stars, obtained a distance modulus of $\mu = 29.32$ mag 
for M74, and  $\mu = 29.50$ mag for the M74 group, confirming previous 
results by \citet{sohn96}, who preferred to use red supergiants. 
As \citet{vinko04} comment, this value was
derived using the Galactic absorption maps of \citet{burstein82}, which give a
value of the total absorption in the $B$ band of $A_B = 0.13$ mag. On the other
hand, the more recent maps published by \citet{schlegel98} predict a larger
absorption, $A_B = 0.301$ mag, leading to a reduced distance, $\mu = 29.15$ mag.

As for the reddening, combining the Galactic extinction of \citet{schlegel98},
$E(B-V)_G = 0.075$ mag, with the small reddening within M74 
\citep[$E(B-V)_H = 0.020$ mag]{tak02}, a total value of $E(B-V) = 0.09$ mag
is used. 

Clearly, the uncertainty associated with the distance is the largest
contribution to the overall uncertainty on the luminosity of SN\,2002ap.
Therefore, in this paper we have chosen to adopt the same values used in previous
modelling and analysis papers 
\citep[($\mu = 29.50$ mag, $E(B-V) = 0.09$ mag)]{maz02,yosh03} in order to make the
results immediately comparable. The effect of adopting a different value for the
distance is discussed below.

\section{One-Zone Models}

The driving parameters for the one-zone fits are the line width, the mass of
\Nifs, and the masses of the other elements. Fitting the widths of the complex
[\FeII] blend near 5200~\AA\ gives a measure of the distribution of \Nifs, since
most Fe is the product of the decay of \Nifs. This blend can be fitted for a
nebular velocity of between 6200 and 5200 \kms. The line width slowly decreases
over the period considered, which is in good agreement with what was found for
SN\,1998bw \citep{maz01}, although in that case the velocities were significantly
larger. This behavior indicates that the outer parts of the ejecta are becoming
progressively more difficult to excite as the density decreases owing to the
expansion. Unlike the case of SN\,1998bw, however, the velocity that is required
to fit the [\FeII] lines is also appropriate for the [\OI] line. This is typical
of lower-energy SNe~Ic \citep[\eg SN\,1994I,][]{sauer06}, and indicates that the
degree of asphericity of the explosion is small. The only apparent deviation from
sphericity is the presence of narrow emission spikes in the cores of [\OI]
$\lambda\lambda$6300, 6363  and \MgI] $\lambda$4571, at velocities below
2000\,\kms \citep[][Figure 17]{fol03}.  Such a feature was also observed in
SN\,2006aj \citep{maz07}. 

The masses of the other elements are determined by fitting the various emission
lines. One caveat is that two intermediate-mass elements that are expected to  be
abundant in the SN ejecta are only diagnosed by a single line. The only line of
sulfur is [\SI] $\lambda\lambda$4069, which is blended with various Fe lines.
Silicon, a highly abundant element, also has no strong or isolated line at optical
wavelenghts. A shoulder on the red side of [\OI] $\lambda\lambda$6300, 6363 may be
due to [\SiI] $\lambda$6527. We have determined the masses of Si and S by
trying to reproduce those two features, but the uncertainty involved is large, affecting
also the estimate of the mass in the ejecta.  In fact, both silicon and sulfur
have strong lines in the infrared ([\SiI] 1.61, 1.65 $\mu$m, [\SI] 1.08, 1.13
$\mu$m), which act as efficient coolants. Thus, increasing the mass of these two
elements requires all other masses, including that of \Nifs, to be increased.
Additionally, since in the later spectra it is not possible to reproduce
accurately the red shoulder of the [\OI] line, there may be some doubt as to the
actual contribution of silicon to that feature. More accurate statements about the
masses of these elements would require the availability of infrared spectra. 

As in SN\,1998bw, the spectra of SN\,2002ap show lines of [\FeII], but not
[\FeIII]. This implies a rather low degree of ionization, which can be
reproduced assuming significant clumping to favor recombination. As in
\citet{maz01}, a volume filling factor of 0.1 was adopted in all models to
achieve this. The models with stratified density and composition discussed in
the next section are useful to address this issue.

The results are summarized in Table 2, and the time series of the spectral fits
is shown in Figures 1 and 2. The electron temperature and density in the nebula
decrease with time. The inferred value of $M$(\Nifs) is $\sim 0.10\, \Msun$.
Small oscillations around this value in the various fits are probably due to a
combination of inconsistent flux calibration and incomplete wavelength
coverage, in particular for spectra that do not include the \MgI] line. Only in
the two very late-time spectra does the \Nifs\ mass increase significantly, but these
spectra are calibrated in flux using an extrapolation of the observed
photometry, making the results less reliable.  

In order to fit the spectra, we
require an ejected mass of $\sim 1.6-1.9\, \Msun$ below a velocity of
5500\,\kms. Of this mass, $\sim 0.7\,\Msun$ is oxygen. The density profile
(CO100/4) used by \citet{maz02} to fit the early-time light curve and spectra
only contained $\sim 0.8\,\Msun$ below the same velocity.  Interestingly, the
larger mass is in good agreement with the results of \citet{mae03}, who used a
two-component, one-dimensional density distribution to reproduce the light
curve of SN\,2002ap and obtained a \Nifs\ mass of $0.08\, \Msun$ and an ejected
mass of $1.6\, \Msun$ below 5750\,\kms\ (see also \citealt{tom05}). Although
the mass at low velocity may be somewhat overestimated by the one-zone models,
which overfit the flux at the lowest velocities where the observed [\OI] line
has a narrow core, the need for additional mass at low velocity is clear. 
Similar results were also obtained for other SNe~Ic
\citep[e.g.,][]{sauer06,maz07}. They suggest the presence of a dense,
oxygen-rich core in the ejecta, which would be most naturally explained as the
result of an aspherical explosion. 

One place where our models fail is the emission near 7700~\AA. This is likely to be
O~I $\lambda$7774. The excitation temperature of this line is significantly
higher than the nebular temperatures obtained in our models, suggesting that the
flux in the line may be caused by recombination or non-thermal excitation by fast
electrons.

Therefore, while the results from the one-zone models generally confirm
findings from the light curve and spectral study of \citet{maz02}, they suggest
that the improvements introduced by \citet{mae03} are realistic, and that a
simple one-dimensional explosion model is not sufficient to explain the
behavior of SN\,2002ap.

\section{Multi-Zone Models}

We have modelled the same 11 spectra of SN\,2002ap with the multi-zone code
described in Section 2. As a model of the explosion, we selected CO100/4, which
was used by \citet{maz02} to fit the early-time light curve and spectra of
SN\,2002ap.

Our multi-zone code is still one-dimensional, but by allowing us to investigate
the element distribution in velocity space at an epoch when the ejecta are
fully transparent, it can give us indirect clues of possible asymmetries
through unusual abundance distributions. Therefore, although we started our
models using the density profile and the abundances of CO100/4, we allowed the
abundances to vary so as to achieve the best possible fit to the data.
Repeating this exercise over several epochs ensures the validity of the
results.

Also, because the ejecta are assumed to be fully transparent, once the abundance
distribution has been defined from fitting one spectrum, it should not need to
be changed to reproduce the other epochs. Ideally, this should be done with the
earliest nebular spectrum, since this is likely to show emission from a broader
range of velocities than later spectra, as discussed in the previous section.

Since the first spectrum of our series does not cover the \MgI] $\lambda$4571
line, which is an important coolant, we used as reference the Lick Observatory 8
June 2002 spectrum, which offers very broad wavelength coverage at a high
signal-to-noise ratio. The model used to fit that spectrum gave good results for
the other spectra, with only small changes in composition, which may be attributed
to the variable observing conditions and flux calibration. 

The series of our synthetic spectra is shown in Figures 4 and 5. In order to
reproduce the width of the [\FeII] lines, the mass fraction of \Nifs\ must be
small ($\sim 1$\%) at $9000 < v < 17,000$\,\kms, and then slowly increase inward,
reaching $\sim 10$\% at $3000 < v < 6000$\,\kms. At the lowest velocities, below
3000\,\kms, the abundance drops again to $\sim 2$\%. For the highest velocities
($v > 17,000$\,\kms), the value of the abundance derived from the early-time
modelling ($\sim 10$\%) was used, but this does not affect the nebular spectra. 
The oxygen mass fraction, on the other hand, does not change significantly with
velocity. It is $\sim 65$\% at $v > 10,000$\,\kms, decreasing to $\sim 40$\% at
$2000< v < 7000$\,\kms, a region that is dominated by \Nifs, Si, and S. The
abundance of S was determined from the weak [\SI] $\lambda$4069 line, and the Si
abundance was assumed to have a constant ratio of 3:1 with respect to that of S.
With this assumption, the emission on the red side of the [\OI] line is not fully
reproduced, suggesting that [\SiII] is not the only line contributing to it, or
that a larger mass, or a larger Si/S ratio should be adopted.  Other elements
behave as expected. The Mg and Ca abundances also increase inward, while C
decreases with O. 

One problem using model CO100/4 is that the strong, narrow cores of [\OI]
$\lambda\lambda$6300, 6363 and \MgI] $\lambda$4571 cannot be reproduced. The 
characteristic velocities of these cores are in fact $< 2000$\,\kms, but 
CO100/4 has a density ``hole'' at $v < 3000$\,\kms.  Therefore, following
\citet{mae03}, we added a dense inner region at $v < 3000$\,\kms. The
composition of this region must be dominated by O, with C, Mg, and other
intermediate-mass elements also present so that the line cores can be
reproduced. However, only little \Fefs\ must be present there, or the [\FeII]
feature would be negatively affected. This inner zone contains only $\sim 0.2\,
\Msun$ of material, mostly oxygen. The total \Nifs\ mass below 3000\,\kms\ is
$\sim 0.01\, \Msun$, in agreement with the result of \citet{mae03}. The oxygen
mass below 5750\,\kms\ in these models is $0.35\, \Msun$, which is
significantly less than in the one-zone models. The total \Nifs\ mass is,
however, slightly larger ($\sim 0.11\, \Msun$). The total ejected mass is
\Mej\,$\approx 2.5\, \Msun$, of which $\sim 1.3\, \Msun$ is oxygen. 

The main results obtained from the multi-zone models are recapped in Table 2.
The last two spectra again require a somewhat larger \Nifs\ mass. Because of
the uncertain flux calibration of these spectra, we do not regard this as an
inconsistency of our modelling procedure. The oxygen mass is, however, much more
stable than in the one-zone models, showing that the multi-zone approach is
more accurate, and that the explosion model that we used is a good
representation of the ejecta of SN\,2002ap in the late phase as well as the early
phase. 

One further point concerns clumping. Introducing a density gradient should
reduce the ionization degree at low velocity. Thus, it may be expected that in
the stratified models clumping would not be necessary. However, even in these
models, if clumping is not used the Fe ionization is too high at all
velocities. A filling factor of 0.1 was therefore used in all zones with $v >
2000$\,\kms --- that is, wherever \Nifs\ is present. No clumping is necessary
in the O-dominated inner zone, below 2000\,\kms. There are a number of possible
explanations for this. One is that clumping does indeed exist. The lack of
direct evidence (\eg narrow emission features at different velocities) suggests
that any clumps must be small and evenly distributed in velocity space. Another
possibility is that \Nifs\ is located mostly near the jet direction in an
axisymmetric explosion. In this case, the actual density of Fe in the
effectively emitting volume in the nebular phase would be higher than what is
suggested by a spherically symmetric approach like the one used here. Clumping
would then be a proxy for this aspherical distribution. The total mass estimate
would not be affected, since in the nebular phase each emitting ion contributes
to the line profile, independent of its position. An axisymmetric
configuration was indeed suggested for the outer part of the ejecta as a
possible interpretation of the detection of significant line polarization in
the early-time spectra of SN\,2002ap \citep{kaw02,leo02}.  If significant
asymmetry affected the inner part of the ejecta, the profiles of the nebular
emission lines would depend sensitively on viewing angle, as demonstrated by
\citet{mae02} for SN\,1998bw. However, in this case the lack of a coincident
GRB and of dramatic signatures of asphericity in the nebular spectra suggest
that any deep asphericity was most likely weak.

\section{The Geometry of the Ejecta of SN\,2002ap}

We have modelled a series of nebular spectra of the broad-lined SN\,2002ap, 
using a one-zone non-LTE code and a multi-zone code where $\gamma$-ray 
deposition and line emission are computed more accurately based on a
pre-defined explosion model. Both simulations yield a good reproduction of the
[\FeII] lines, but the narrow cores observed in [\OI] $\lambda\lambda$6300,
6363 can only be reproduced assuming the presence of a dense, O-dominated inner
zone. In the one-zone synthetic spectra, a mass excess of $\sim 0.7\, \Msun$ is
needed to reproduce qualitatively the [\OI] emission line.  In the multi-zone
model, which reproduces the line profiles much more accurately, the mass excess
is smaller, $\sim 0.2 \Msun$, but the presence of oxygen-dominated material
moving at low velocity is confirmed.  This material is not predicted by
one-dimensional explosion models, but its presence was suggested in various
hypernovae based on light-curve studies \citep{mae03}. 

Similar results were also obtained for the normal SN\,Ic 1994I \citep{sauer06} as
well as for the XRF/SN 2006aj \citep{maz07}. SN\,1994I shows emission in the [\OI]
$\lambda\lambda$6300, 6363 line at velocities below $2000$\,\kms\ \citep{fil95},
while the  corresponding one-dimensional explosion model predicts the absence of
material at low velocities. The rounded profile suggests a smooth distribution of
oxygen, which does not by itself indicate an aspherical explosion, but the very
presence of material does, and it can be modelled adding $0.2 \Msun$ of
oxygen-dominated mass at low velocity. SN\,2006aj shows a peak in [\OI]
$\lambda\lambda$6300, 6363 at velocities below 2000\,\kms, although not quite as
sharp as in SN\,2002ap (Figure 4). This was modelled for a mass excess of $\sim
0.7 \Msun$ over the one-dimensional explosion model \citep{maz07}. The presence of
a dense inner zone in all SNe~Ic suggests that these explosions are intrinsically
aspherical, as is also deduced from polarization studies
\citep{wang01,leo02,fil04,leo05}.

Figure~6 shows a comparison of the [\OI] $\lambda\lambda$6300, 6363 line in
SN\,2002ap, and in SNe\,2006aj, 1998bw, and 2003jd (Mazzali et al. 2007, 2001,
and 2005, respectively).  The line in SN\,2006aj has a full-width at
half-maximum intensity (FWHM) of $\sim 8000$\,\kms, while for the other three
SNe the FWHM is $\sim 6000$\,\kms. The line profile in SN\,2006aj is not very
sharp, indicating that the nebula is to a good approximation spherically
symmetric \citep{maz07}. In this case the [\OI] line velocity should correspond
to the real expansion velocity. The sharp profile in SN\,1998bw was interpreted
as a disk-like distribution of oxygen viewed from a near-polar direction.
Although the low observed velocity is partly the result of a projection effect,
the actual velocity of the oxygen ($\sim 8000$\,\kms) is smaller than the
velocity of Fe \citep{mae02}. The [\OI] line of SN\,2003jd has a width similar
to that of SN\,1998bw, but the double-peaked profile suggests that we are
viewing the oxygen-rich disk from near its plane \citep{maz05}. The case of
SN\,2002ap seems intermediate between those of SNe~1998bw and 2006aj. The line
can be separated into two components. A broad base, of width similar to that in
SNe 1998bw and 2003jd, is present down to velocities of $\sim 3000$\,\kms\ and
is well fitted by a one-zone model. Its profile is reminiscent of that of
SN\,2003jd, but the presence of the sharp core makes it impossible to
distinguish between a flat-topped profile, characteristic of a shell-like
spherical distribution, and a double-peaked profile as in a disk viewed
edge-on. This core is narrower than even the corresponding profile of
SN\,1998bw. 

There may be two explanations for this type of profile. One is that in
SN\,2002ap the outer part of the oxygen, from $v \approx 3000$ out to $v
\approx 8000$\,\kms, has a distribution that is not far from spherical, giving
rise to a flat-topped profile. Below $\sim 3000$\,\kms, on the other hand, we
may be seeing a situation qualitatively similar to that of SN\,1998bw, with the
oxygen distributed in a disk-like region oriented not far from face-on.  The
size and mass of this disk-like region are, however, much smaller in SN\,2002ap
than in SN\,1998bw. The other possibility is that the outer part reflects a
disk-like distribution of most of the oxygen viewed not far from its plane,
like in SN\,2003jd. The broad component of the profile would then be
double-peaked rather than flat-topped, and the inner component would contain
more mass, so that its emission could fill the valley between the two peaks.
The sharpness of this inner-component emission may again indicate a disk-like
distribution viewed face-on, but in this case two main orientations would exist
in the ejecta, which may not be easy to explain in terms of the collapse
mechanism. Alternatively, the inner region could simply be sharply peaked in
density, as reproduced by our multi-zone model, but not grossly aspherical.

In both scenarios the lack of a GRB may be justified. If the inner explosion was
aspherical, but the asphericity did not propagate to the intermediate part of the
ejecta (3000 to $8000$\,\kms), it is unlikely that a beam of material moving at
relativistic velocities was produced, yet the overenergetic explosion could have
caused enough material to move at high velocities ($\sim 30000$\,\kms), giving
rise to the broad absorption features in the early spectra. The distribution of
this material may not have been very far from spherical: polarization studies
\citep{kaw02,leo02} suggest a degree of asphericity of $\sim 10$\%. On the other hand,
if the outer explosion was aspherical, a relativistic jet may have been produced,
but it would not have been pointing to us. This latter possibility is, however,
not supported by the lack of an X-ray detection \citep{soria03} and the weakness
of the radio signal \citep{berger02}, so we tend to favor an explosion that was
not far from  spherical, except in the innermost part.

Both simulations agree that the \Nifs\ mass ejected by SN\,2002ap is $\sim 0.10\,
\Msun$, a slightly larger value than that predicted by studies of the early-time light
curve \citep{maz02}, but in agreement with the late-time light curve
\citep{mae03,tom05}. The multi-zone model confirms that only a small fraction of
this ($\sim 0.01\, \Msun$) is located in the innermost zone, a clear indication
that the explosion was asymmetric to some degree.

Other global values that can be derived rather accurately from a
one-dimensional study of the nebular spectra ($M$(\Nifs)\,$\approx 0.11\,
\Msun$, \Mej\,$\approx 2.5\, \Msun$) are small compared to those of other
broad-lined SNe, confirming previous results based on the early-time light
curve and spectra.  These values point to a progenitor less massive than for
objects such as SN 1998bw or SN 2003dh, although still a massive star ($M
\approx 22-25\, \Msun$) that could have produced a black hole. This also
suggests that the ``hypernova'' properties of SN\,2002ap were less extreme,
including asymmetry and the launching of a GRB.

The estimated mass of \Nifs\ would be smaller if we adopted a shorter distance.
If the distance modulus was $\mu = 29.15$ mag, the \Nifs\ mass would be reduced by
$\sim 15$\%. The mass and energy of the ejecta would also be similarly reduced,
but the change is sufficiently small that the shape of the light curve would
not be affected. 

Indirect indications of asymmetry can be obtained in our one-dimensional study, 
but the details of the exact properties are not.  The development of axisymmetric
explosion models suitable for SN\,2002ap, and the study of the nebular spectra
using three-dimensional spectrum synthesis codes as in \citet{mae02}, are
currently planned. 

\bigskip

The work of A.V.F.'s group at UC Berkeley is supported by National
Science Foundation grant AST--0607485. We thank the Subaru and Lick 
Observatory staffs for their assistance with the observations.

%%%%%%%%%%%%%%%%%%%%%%%%%%%%%%%%%%%%%%%%%%%%%%%%%%%%%%%%%%%%%%%%%%%%%%%%%%%%

%%%%%%%%%%%%%%%%%%%%%%%%%%%%%%%% FIGURES %%%%%%%%%%%%%%%%%%%%%%%%%%%%%%%%%%%%%

%%%%%%%%%%%%%%%%%%%%%%%%%%%%%%%% FIGURE 1 %%%%%%%%%%%%%%%%%%%%%%%%%%%%%%%%%%%%%

\begin{figure} 
%\plotone{neb02ap_newspec_f1.eps}
\plotone{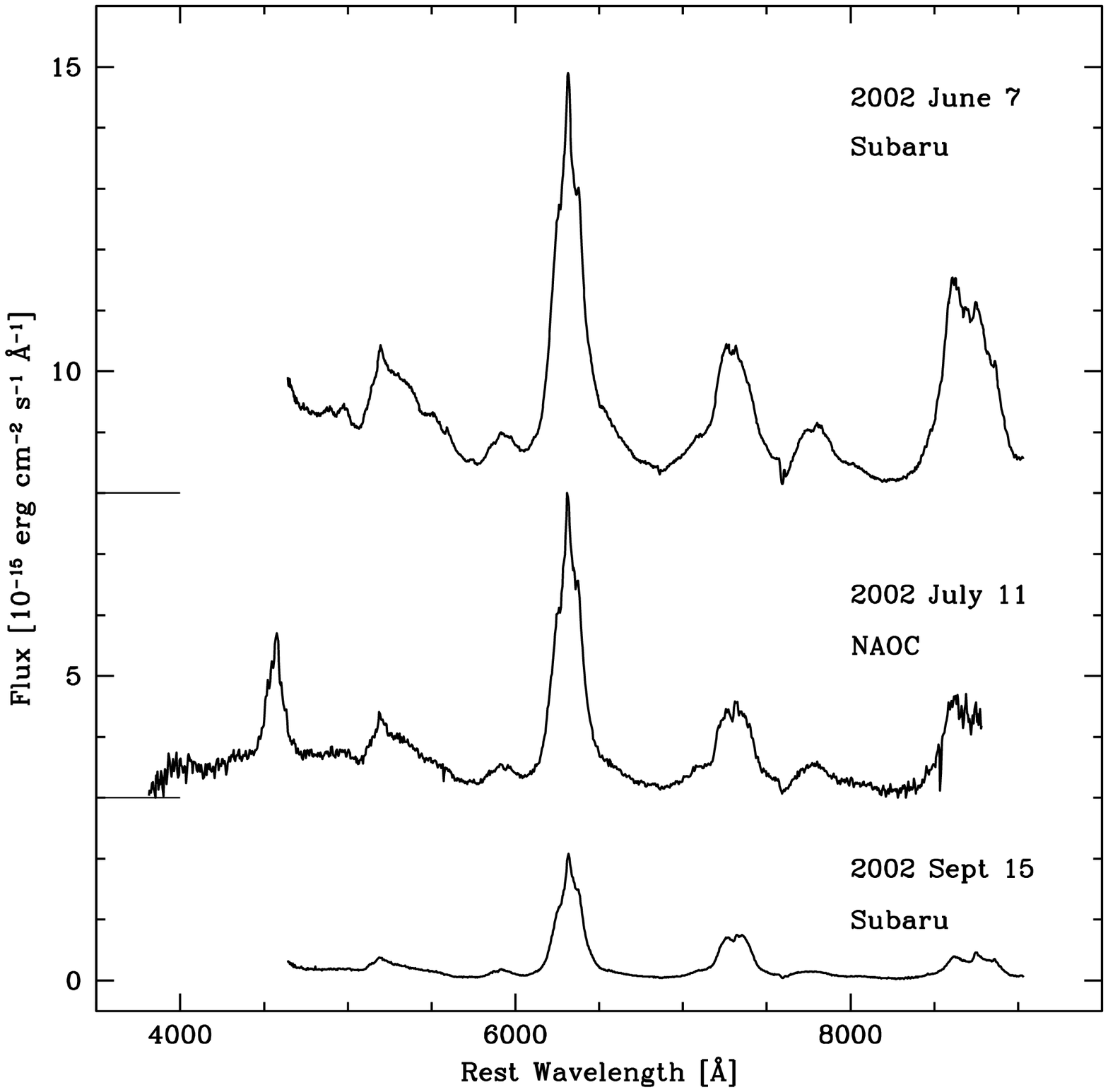}
\figcaption[]{The nebular spectra of SN\,2002ap obtained at Subaru and NAOC. The
two earlier spectra have been shifted upward by adding a constant. The zero
level of the flux is shown by a thin horizontal line.}
\end{figure}

%%%%%%%%%%%%%%%%%%%%%%%%%%%%%%%% FIGURE 2 %%%%%%%%%%%%%%%%%%%%%%%%%%%%%%%%%%%%%

\begin{figure} 
\epsscale{0.90}
%\plotone{neb02ap_all_1z_1_f2.eps}
\plotone{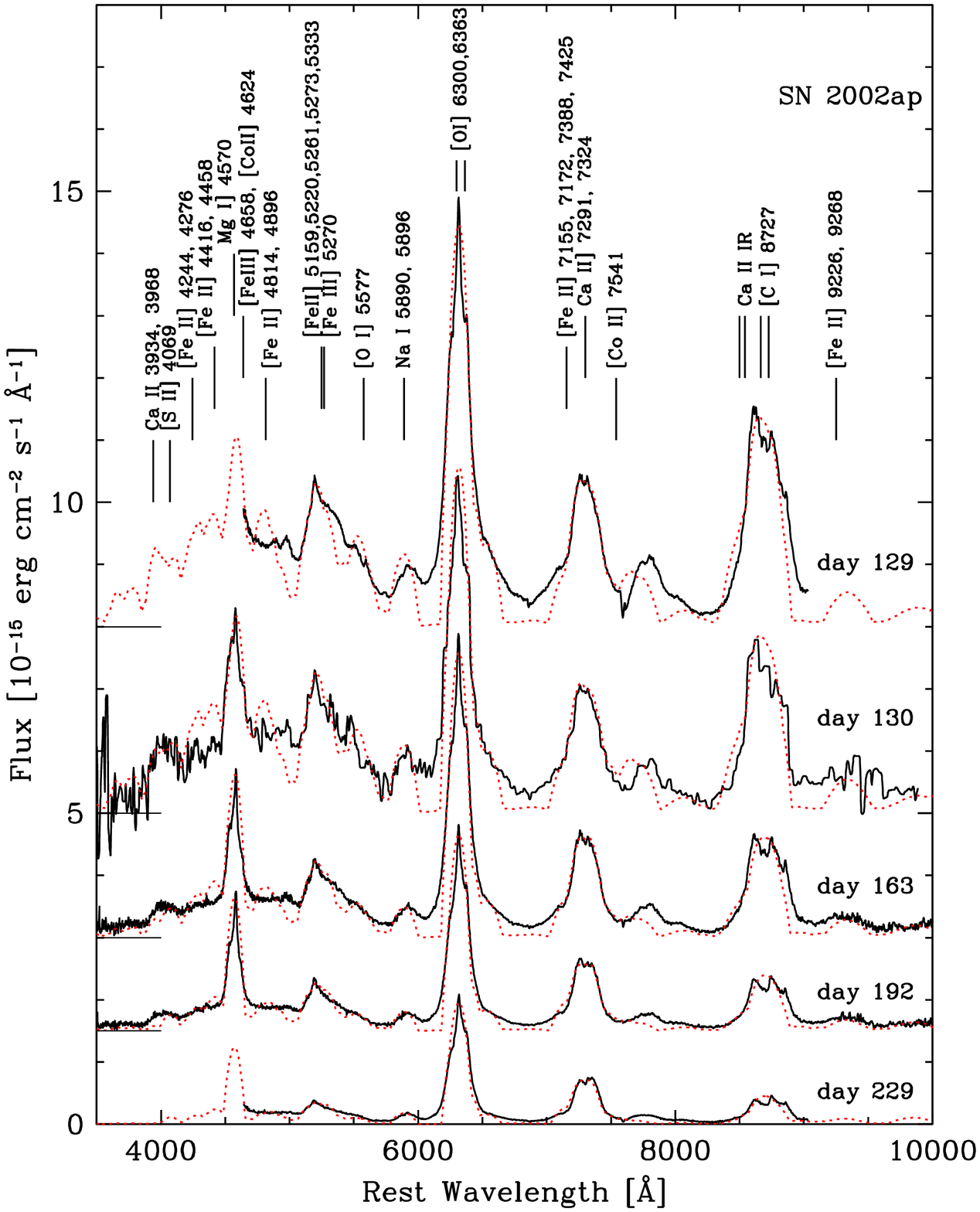}
\figcaption[]{The nebular spectra of SN\,2002ap obtained from 2002 June to 2002
September, compared to the synthetic spectra computed with the one-zone model 
(dotted/red). Except for the lowest one, 
all spectra have been shifted upward by arbitrary amounts. Long tick marks on 
the left ordinate axis show the zero flux level for each spectrum. [{\em See 
the electronic edition for a color version of this figure.}]}
\end{figure}

%%%%%%%%%%%%%%%%%%%%%%%%%%%%%%%% FIGURE 3 %%%%%%%%%%%%%%%%%%%%%%%%%%%%%%%%%%%%%

\begin{figure} 
\epsscale{0.90}
%\plotone{neb02ap_all_1z_2_f3.eps}
\plotone{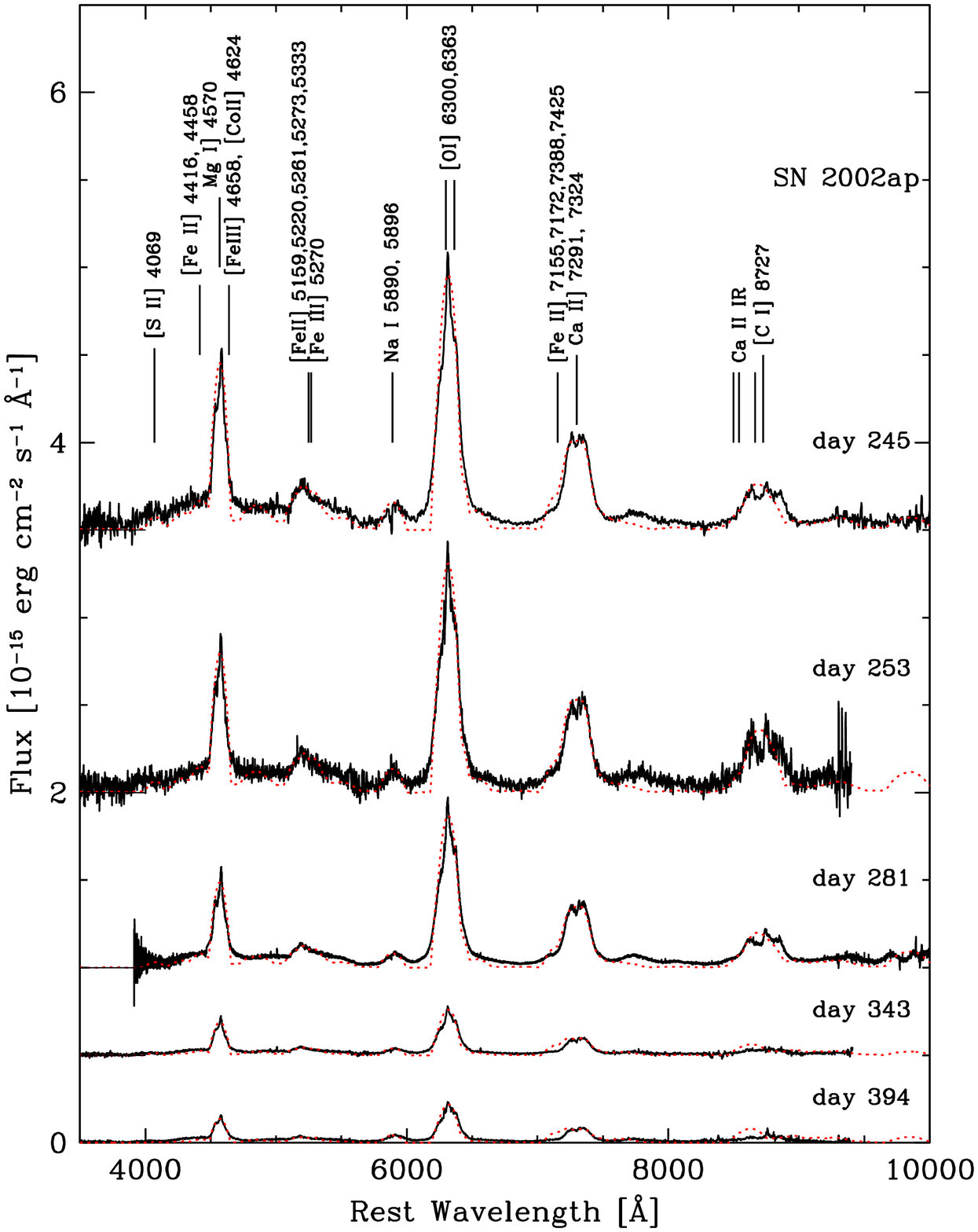}
\figcaption[]{The nebular spectra of SN\,2002ap obtained from 2002 October to 
2003 February, compared to the synthetic spectra computed with the one-zone 
model (dotted/red). Except for the lowest one, all spectra have been shifted 
upward by arbitrary amounts. Both the lowest spectrum and the corresponding 
model have been multiplied by a factor of 2 for display purposes. 
Long tick marks on 
the left ordinate axis show the zero flux level for each spectrum. [{\em See 
the electronic edition for a color version of this figure.}]}
\end{figure}

%%%%%%%%%%%%%%%%%%%%%%%%%%%%%%%% FIGURE 4 %%%%%%%%%%%%%%%%%%%%%%%%%%%%%%%%%%%%%

\begin{figure}
\epsscale{0.90}
%\plotone{neb02ap_all_sh_1_f4.eps}
\plotone{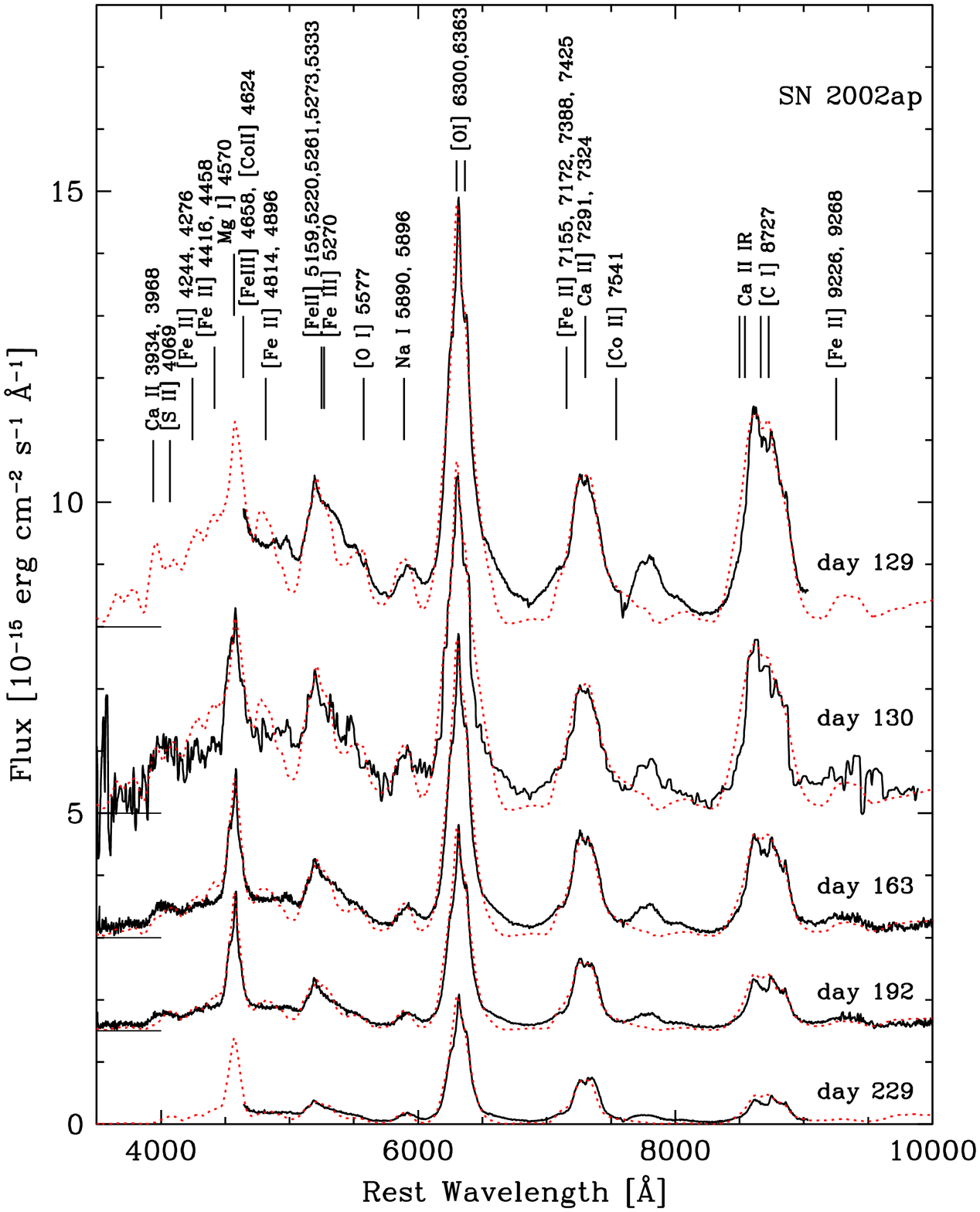}
\figcaption[]{The nebular spectra of SN\,2002ap obtained from 2002 June to 2002
September, compared to the synthetic spectra obtained with the multi-shell 
model (dotted/red). Except for the lowest one, 
all spectra have been shifted upward by arbitrary amounts. Long tick marks on 
the left ordinate axis show the zero flux level for each spectrum. [{\em See 
the electronic edition for a color version of this figure.}]}
\end{figure}

%%%%%%%%%%%%%%%%%%%%%%%%%%%%%%%% FIGURE 5 %%%%%%%%%%%%%%%%%%%%%%%%%%%%%%%%%%%%%

\begin{figure} 
\epsscale{0.90}
%\plotone{neb02ap_all_sh_2_f5.eps}
\plotone{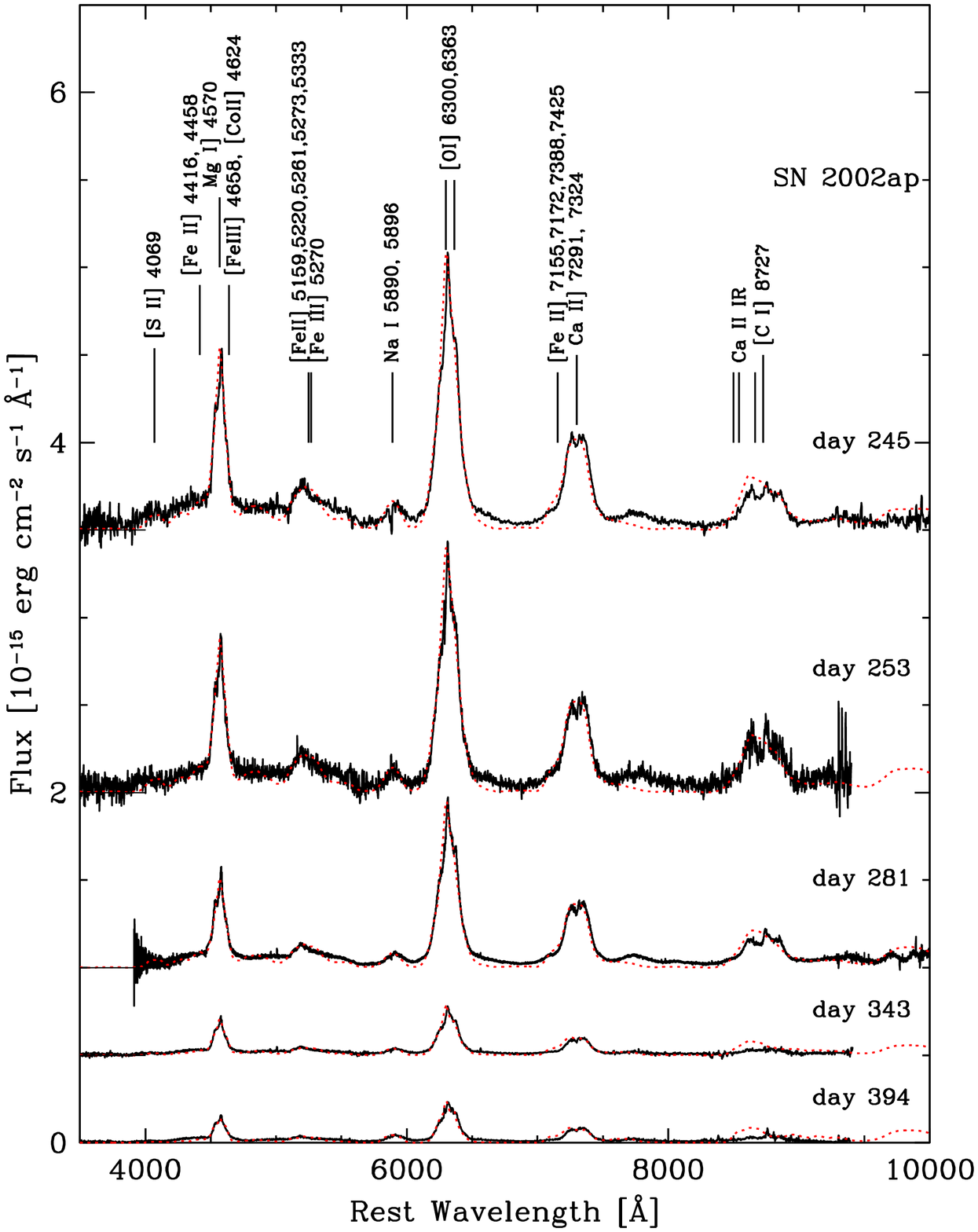}
\figcaption[]{The nebular spectra of SN\,2002ap obtained from 2002 October to 
2003 February, compared to the synthetic spectra computed with the multi-shell  
model (dotted/red). Except for the lowest one, all spectra have been shifted 
upward by arbitrary amounts. Both the lowest spectrum and the corresponding 
model have been multiplied by a factor of 2 for display purposes. 
Long tick marks on 
the left ordinate axis show the zero flux level for each spectrum. [{\em See 
the electronic edition for a color version of this figure.}]}
\end{figure}

%%%%%%%%%%%%%%%%%%%%%%%%%%%%%%%% FIGURE 6 %%%%%%%%%%%%%%%%%%%%%%%%%%%%%%%%%%%%%

\begin{figure}
\epsscale{0.90}
%\plotone{sne02ap98bw06ajOI_f6.eps}
\plotone{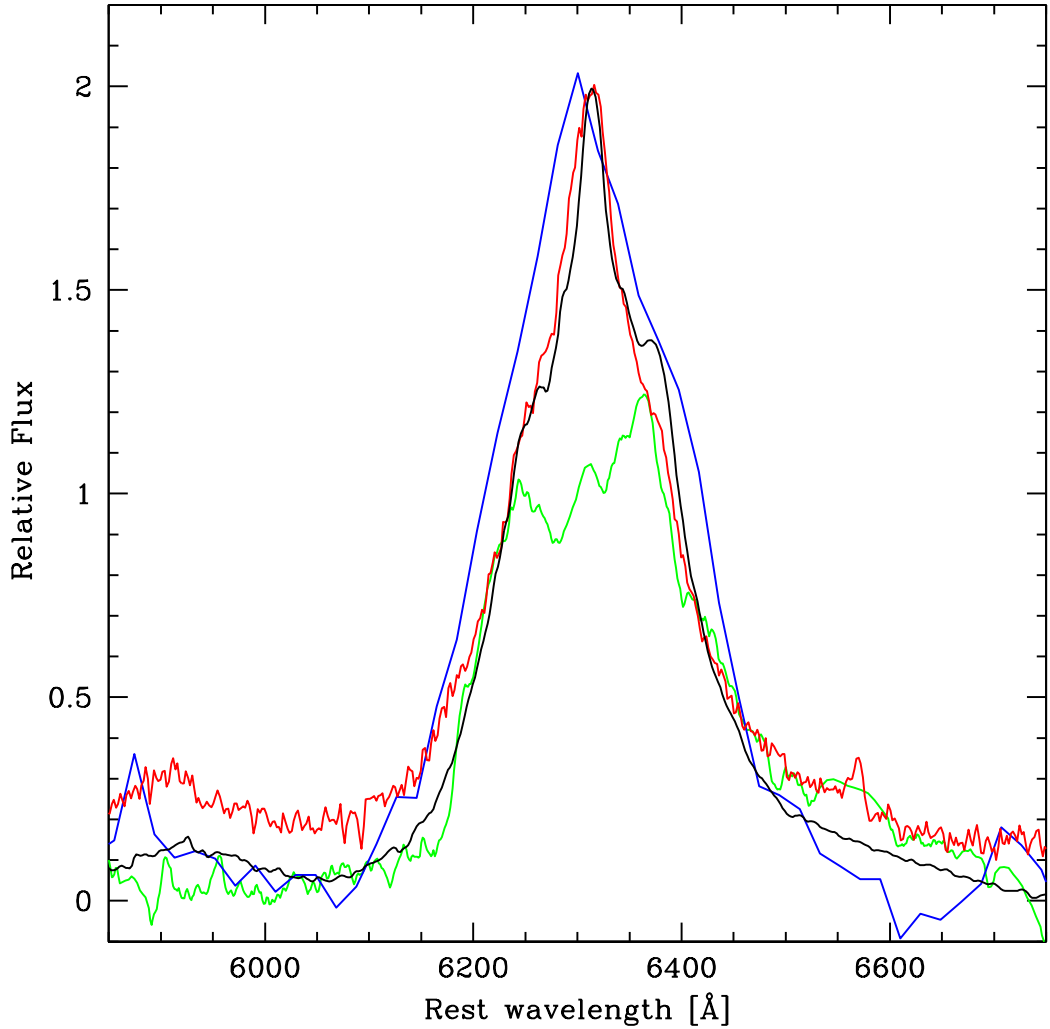}
\figcaption[]{A comparison of the [\OI] $\lambda\lambda$6300, 6363 line of 
SNe 2002ap (solid grey/black line), 2006aj (long dashed/blue line), 
1998bw \citep[dotted/red line,][]{pat01}, and 
2003jd \citep[short dashed/green line,][]{maz05}. 
%The profile of the line in 
%SN\,2006aj is less peaked than that of SN\,1998bw, although the average 
%expansion velocity of other elements is larger in SN\,1998bw than in 
%SN\,2006aj, indicating a smaller degree of asphericity in SN\,2006aj.
Spectra have been scaled by various amounts to facilitate comparison.  
[{\em See the electronic edition for a color version of this figure.}]}
\end{figure}

%%%%%%%%%%%%%%%%%%%%%%%%%%%%%%%%%%%%%%%%%%%%%%%%%%%%%%%%%%%%%%%%%%%%%%%%%%%%%

%%%%%%%%%%%%%%%%%%%%%%%%%%%%%%%% TABLES %%%%%%%%%%%%%%%%%%%%%%%%%%%%%%%%%%%%%%%

%%%%%%%%%%%%%%%%%%%%%%%% Table 1 %%%%%%%%%%%%%%%%%%%%%%%%%%%%%%%%%%%%%%%

\begin{deluxetable}{rccccccccc}
\tablewidth{0pt}
\tabletypesize{\scriptsize}
\tablenum{1}
\tablecaption{Parameters of the One-Zone Synthetic Spectra}
\tablehead{\colhead{UT Date} &
\colhead{SN epoch} &
\colhead{Telescope} &
\colhead{velocity} &
\colhead{$M$(\Nifs)} &
\colhead{$M$(O)} &
\colhead{$M$(Si)} &
\colhead{$M$(tot)} &
\colhead{$T_e$} &
\colhead{log ($n_e$/}   \\
\colhead{~} &
\colhead{[days]\tablenotemark{a}} &
\colhead{~} &
\colhead{[km s$^{-1}$]} &
\colhead{[$\Msun$]} &
\colhead{[$\Msun$]} &
\colhead{[$\Msun$]} &
\colhead{[$\Msun$]} &
\colhead{[K]} &
\colhead{(g~cm$^{-3}$))}  }
\startdata
 7 June 2002 & 129 &  Subaru  & 6200 & 0.090 & 0.71 & 0.60 & 1.78 & 5511 & 7.58 \\
 8 June 2002 & 130 & Lick 3~m & 6200 & 0.092 & 0.62 & 0.50 & 1.58 & 5488 & 7.56 \\
11 July 2002 & 163 & Lick 3~m & 5750 & 0.088 & 0.71 & 0.57 & 1.73 & 5051 & 7.34 \\
11 July 2002 & 163 &   NAOC   & 5750 & 0.090 & 0.71 & 0.58 & 1.73 & 5082 & 7.35 \\
 9  Aug 2002 & 192 & Lick 3~m & 5550 & 0.093 & 0.66 & 0.60 & 1.69 & 4751 & 7.17 \\
15 Sept 2002 & 229 &  Subaru  & 5550 & 0.098 & 0.70 & 0.74 & 1.90 & 4360 & 6.95 \\
 1  Oct 2002 & 245 & Lick 3~m & 5550 & 0.102 & 0.67 & 0.78 & 1.86 & 4209 & 6.86 \\
 8  Oct 2002 & 253 & Lick 3~m & 5550 & 0.103 & 0.66 & 0.70 & 1.88 & 4160 & 6.82 \\
 6  Nov 2002 & 281 & Lick 3~m & 5550 & 0.110 & 0.71 & 0.80 & 2.04 & 3903 & 6.69 \\
 7  Jan 2003 & 343 & Lick 3~m & 5450 & 0.145 & 0.61 & 0.80 & 1.93 & 3422 & 6.49 \\
27  Feb 2003 & 394 & Lick 3~m & 5200 & 0.170 & 0.62 & 0.80 & 1.92 & 3092 & 6.37 \\
\enddata
\tablenotetext{a}{The epoch is given from the putative date of explosion, 29 Jan
2002 \citep{maz02}.}
\end{deluxetable}
%%%%%%%%%%%%%%%%%%%%%%%%%%%%%%%%%%%%%%%%%%%%%%%%%%%%%%%%%%%%%%%%%%%%

%%%%%%%%%%%%%%%%%%%%%%%% Table 2 %%%%%%%%%%%%%%%%%%%%%%%%%%%%%%%%%%%%%%%

\begin{deluxetable}{rccccc}
\tablewidth{0pt}
\tabletypesize{\scriptsize}
\tablenum{2}
\tablecaption{Parameters of the Multi-Zone Synthetic Spectra}
\tablehead{\colhead{UT Date} &
\colhead{SN epoch} &
\colhead{Telescope} &
\colhead{$M$(\Nifs)} &
\colhead{$M$(O)}   &
\colhead{$M$(Si)}    \\
\colhead{~} &
\colhead{[days]\tablenotemark{a}} &
\colhead{~} &
\colhead{[$\Msun$]} &
\colhead{[$\Msun$]} &
\colhead{[$\Msun$]}   }
\startdata
 7 June 2002 & 129 &  Subaru  &  0.11 & 1.21 & 0.44 \\
 8 June 2002 & 130 & Lick 3~m &  0.11 & 1.13 & 0.58 \\
11 July 2002 & 163 & Lick 3~m &  0.11 & 1.33 & 0.31 \\
11 July 2002 & 163 &   NAOC   &  0.11 & 1.33 & 0.31 \\
 9  Aug 2002 & 192 & Lick 3~m &  0.12 & 1.28 & 0.36 \\
15 Sept 2002 & 229 &  Subaru  &  0.11 & 1.34 & 0.29 \\
 1  Oct 2002 & 245 & Lick 3~m &  0.11 & 1.29 & 0.36 \\
 8  Oct 2002 & 253 & Lick 3~m &  0.11 & 1.29 & 0.33 \\ 
 6  Nov 2002 & 281 & Lick 3~m &  0.11 & 1.29 & 0.31 \\ 
 7  Jan 2003 & 343 & Lick 3~m &  0.13 & 1.28 & 0.36 \\ 
27  Feb 2003 & 394 & Lick 3~m &  0.13 & 1.35 & 0.30 \\ 
\enddata
\tablenotetext{a}{The epoch is given from the putative date of explosion, 29 Jan
2002 \citep{maz02}.}
\end{deluxetable}
%%%%%%%%%%%%%%%%%%%%%%%%%%%%%%%%%%%%%%%%%%%%%%%%%%%%%%%%%%%%%%%%%%%%

\end{document}